\documentclass[aps,twocolumn,pre,showpacs,superscriptaddress,amsmath,amssymb]{revtex4-1}
\usepackage{graphicx}
\usepackage{epstopdf}
\usepackage{amsmath}
\usepackage{amssymb}
\usepackage{color}
\usepackage{txfonts}
\usepackage{verbatim}
\usepackage{cprotect}
\usepackage{hyperref}
\usepackage{CJK}

\newcommand{\+}[1]{\ensuremath{\mathbf{#1}}}

\definecolor{dkgreen}{rgb}{0,0.6,0}
\definecolor{gray}{rgb}{0.5,0.5,0.5}
\definecolor{mauve}{rgb}{0.58,0,0.82}
\definecolor{dkblue}{rgb}{0.19,0.51,0.74}

\begin{document}

\title{Concurrency and reachability in tree-like temporal networks}

\author{Eun Lee}
\affiliation{Carolina Center for Interdisciplinary Applied Mathematics, Department of Mathematics, University of North Carolina at Chapel Hill, USA}
\author{Scott Emmons}
\affiliation{Carolina Center for Interdisciplinary Applied Mathematics, Department of Mathematics, University of North Carolina at Chapel Hill, USA}
\author{Ryan Gibson}
\affiliation{Carolina Center for Interdisciplinary Applied Mathematics, Department of Mathematics, University of North Carolina at Chapel Hill, USA}
\author{James Moody}
\affiliation{Duke Network Analysis Center and Department of Sociology, Duke University, Durham, NC, USA}
\author{Peter J. Mucha$^{1,}$}
\affiliation{Department of Applied Physical Sciences, University of North Carolina at Chapel Hill, USA}

\begin{abstract}
Network properties govern the rate and extent of various spreading processes, from simple contagions to complex cascades. Recently, the analysis of spreading processes has been extended from static networks to temporal networks, where nodes and links appear and disappear. We focus on the effects of ``accessibility", whether there is a temporally consistent path from one node to another, and ``reachability", the density of the corresponding ``accessibility graph" representation of the temporal network. The level of reachability thus inherently limits the possible extent of any spreading process on the temporal network. We study reachability in terms of the overall levels of temporal concurrency between edges and the structural cohesion of the network agglomerating over all edges. We use simulation results and develop heterogeneous mean field model predictions for random networks to better quantify how the properties of the underlying temporal network regulate reachability.
\end{abstract}


\maketitle

\section{Introduction}
\label{sec:introduction}

Social networks are woven together by temporal contacts organized according to various structural details which together form the substrate of infectious dynamics, determining the impacts of spreading diseases and viral information flows. 
Compared to the extensive literature modeling spreading dynamics on static networks, we lack a thorough understanding of the particular effects of the temporal properties of network contacts.
Meanwhile, recent studies have found diverse temporal contact features --- such as distributions of inter-contact times, temporal correlation in inter-contact times, and birth and death of nodes and links --- may have very different impacts on the dynamics of spreading processes~\cite{Barabasi2007,Karsai2011,Goh2011,Naoki2013,Takayuki2018,Holme2014,Lambiotte2016,Colman2016}.

\textit{Concurrency}, broadly defined as `relationships that overlap in time'~\cite{Moody2016}, is one of the key elements affecting the extent and speed of disease spreading. 
Concurrency is a longstanding concept in epidemiology and has been considered in diverse contexts, including for understanding the epidemic potential of HIV/AIDS~\cite{MORRIS1995, WATTS1992, GURSKI2016, Moody2016}. 
Some studies have applied the concept as a property proportional to an average contact rate or an average degree in unit time~\cite{WATTS1992,GURSKI2016} or as a link density of a reachable network converted from a pair-to-pair contact patterns~\cite{MORRIS1995}. 
In another recent study, concurrency was considered as the number of links of an individual in unit time within a generative temporal activity model~\cite{Naoki2017}. 
Whatever the particular definition of concurrency, the general idea in application is that increased concurrency increases the density of the effective network structure over which an infection is transmitted, resulting in a larger number of alternative paths between nodes, thus increasing the potential for greater spread through the population. As such, the general conclusion that higher concurrency increases the potential for epidemic spread seems to be trivial. However, the detailed mechanism of this increase is important to understand and to quantify to assess the impact of concurrency in a particular temporal network setting.

Motivated by previous work~\cite{Moody2016}, we consider the \textit{reachability} of the temporal contact network over which transmission can occur. Reachability is the density of the \emph{accessibility graph} that includes an edge from node $i$ to node $j$ if and only if there is a temporally consistent path originating at $i$ that can reach $j$ in the underlying temporal contact network. 
That is, reachability quantifies the maximum possible impact of the infectious spreading by quantifying the fraction of node pairs that can be accessible via temporally consistent paths (see, e.g., \cite{Grindrod2011,Hartmut2013,Holme2015,Moody2016}). Reachability is a useful metric not only because it measures the maximal substrate of infectious spreading, but also because it indicates how much temporal continuity can be ignored when one uses an aggregated static network to analyze infection dynamics~\cite{Hartmut2013}.

To separate out the influences of the temporal and structural details, we focus as in~\cite{Moody2016} on two critical properties of the temporal network: \textit{temporal concurrency} and \textit{structural cohesion}. 
Temporal concurrency is defined here as the fraction of pairs of links that overlap in time.
Meanwhile, structural cohesion measures the effective connectedness in the underlying topology of the time-aggregated network, ignoring the temporal details of the individual edges. A good measure of structural cohesion should embody the notion that highly cohesive networks should be difficult to separate (i.e., by node or edge removal) into separate components. As such, we employ the definition of structural cohesion as the average number of node-independent paths between two nodes~\cite{Moody2003}, as applied to the network that includes all edges over the total time period studied. We emphasize that structural cohesion is more than a simple function of edge density; rather, it is influenced by the organizational patterns of the connections. In particular, one can observe different amounts of structural cohesion even while keeping the total link density constant. 

By deliberately separating the structural cohesion measurement from that for temporal concurrency, we explore the role of each and the interplay between them in affecting reachability. 
Pairing numerical calculations with an approximate model we develop here, we examine the roles of  these temporal network properties, observing in particular how structural cohesion directly affects the desciption of the use of detours to find temporally-consistent paths between node pairs.
Our approximate model focuses on networks that are tree-like in the sense of having low structural cohesion, in an effort to develop and assess the accuracy of model approximations for the level of reachability in random temporal networks. (We refer the interested reader to Melnik \emph{et al}.~\cite{Melnik2011PRE} for further discussion of what it might mean for a network to be ``tree-like" in this sense.)

We start with detailed definitions of temporal concurrency and structural cohesion in Sec.~\ref{subsec:coh_conc}, continue to describe the methods for constructing our synthetic and sampled empirical networks in Sec.~\ref{subsec:exp_conc} and Sec.~\ref{subsec:construc_network}, respectively, and provide specific quantitative details for numerically computing reachability in Sec.~\ref{subsec:numeric_reach}. We then develop our model approximation for reachability in  Sec.~\ref{sec:model_reach}. 
In Sec.~\ref{sec:results}, we compare numerical measurements and the approximation for reachability on synthetic trees, Erd\H{o}s-R\'enyi networks, and configuration model realizations with exponential degree distributions, before continuing on to the empirical examples studied previously in \cite{Moody2016}. We conclude with a discussion in Sec.~\ref{sec:conclusion} about the effect of temporal concurrency on the reachability and limits of the presently-developed approximation, along with possible future directions for improvement.

\section{Method}
\label{sec:method}
\subsection{Structural cohesion and temporal concurrency}
\label{subsec:coh_conc}

The ease with which disease spreads on a network is typically increased in the presence of multiple diverse alternative paths between nodes. In a temporal network with many links overlapping in time, the  concurrency increases the number of such paths that are temporally consistent, possibly accelerating the spread and increasing the total outbreak size even without increasing the number of contacts in the network.
To study the role of the structural and temporal connectedness, we separately consider the impacts of the \textit{structural cohesion} and \textit{temporal concurrency}, following the approach in~\cite{Moody2016} (summarized above and presented in detail below).

\begin{figure}
\centering
\includegraphics[width=0.98\linewidth]{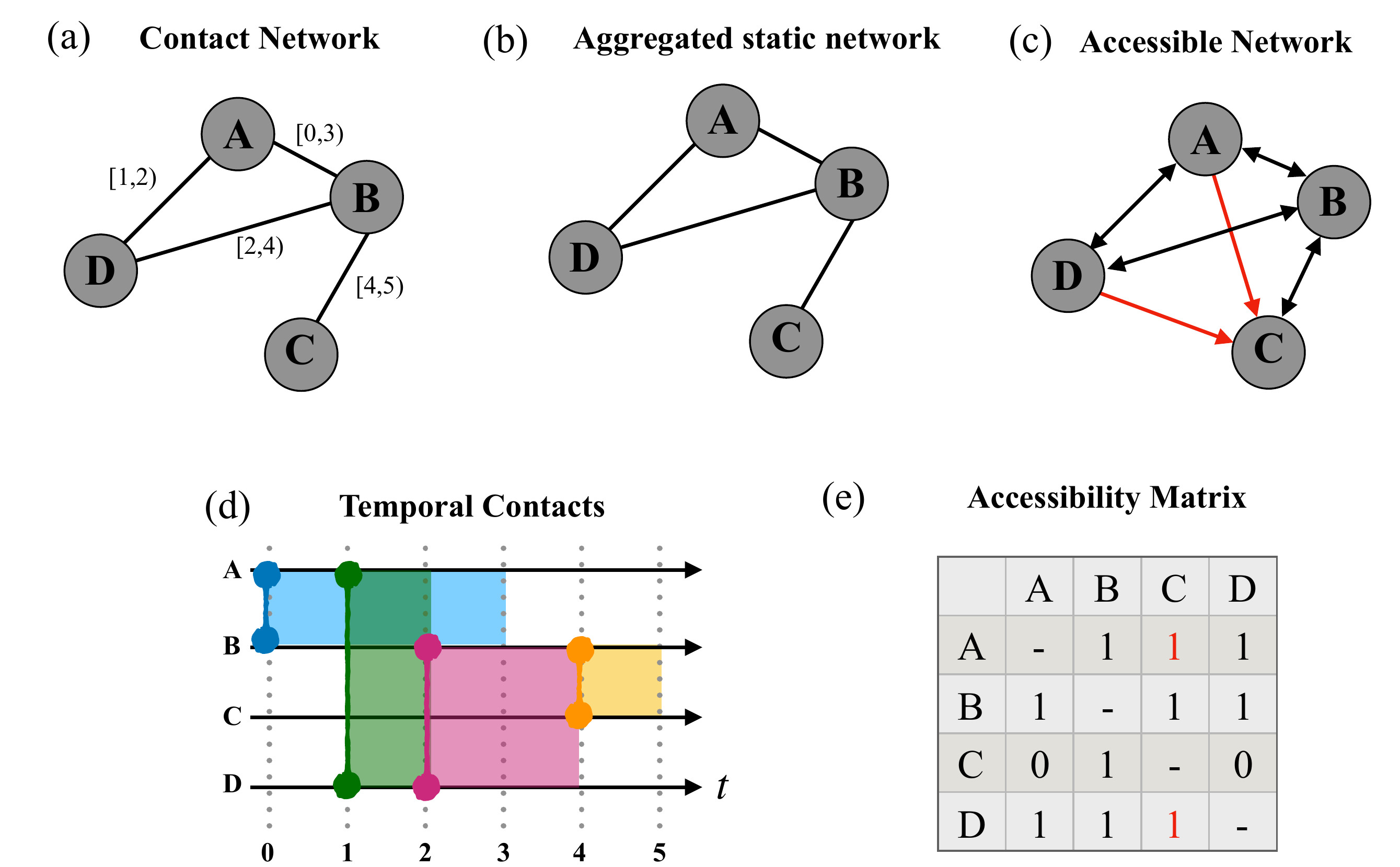}
\caption{Schematics for establishing accessibility. (a) Each edge in the network of nodes $\{{\rm A,B,C,D}\}$ is denoted by a start and end time, e.g., the contact between A and D starts at $t=1$ and continues for $1$ time units to $t=2$. (b) The static network representation aggregating temporal information. (c) The corresponding directed graph of accessibility, demonstrating that asymmetric accessibilities (red arrows) are possible. (d) The contacts represented along a time axis. (e) The accessibility matrix between the four nodes, with the $(i,j)$ element indicates if $j$ is reachable from $i$.}
\label{fig:schem}
\end{figure}

We emphasize that throughout this study we distinctively refer to three related network representations describing the pattern of interaction: (1) the full temporal contact network (Fig.~\ref{fig:schem}(a)), which we assume is undirected; (2) the static aggregated network that includes all links that ever appear (Fig.~\ref{fig:schem}(b)); and (3) the directed accessibility graph (Fig.~\ref{fig:schem}(c)) that describes the presence of temporally-consistent paths between ordered pairs of nodes.

To measure the structural cohesion, $\langle{\kappa}\rangle$, we consider only the static aggregated network representation including all edges that are ever present in the specified temporal network. Within the aggregated network, we seek the number of node-independent paths, $\kappa(i,j)$, available between nodes $i$ and $j$. We employ the shortest path approximation of~\cite{Newman2001} to numerically calculate $\kappa(i,j)$ and then average over all pairs of nodes: 
\begin{equation}
\label{eq:K}
    \langle {\kappa} \rangle=\dfrac{1}{N(N-1)}\sum_{i\neq j}{\kappa(i,j)},
\end{equation}
where $N$ is the size of the network. 

To measure temporal concurrency, $C$, we consider here the single-interval case where the link between nodes $i$ and $j$ (if present) has a single, specified starting time $s(i,j)$ and persists for duration $d(i,j)$. For simplicity, we will assume that start times and durations are each independent and identically distributed (IID) across the edges that ever appear in the aggregated network during the selected total time period, $T$. That is, in particular, the timings of the edges emanating from a given node are necessarily independent of each other. As such, the temporal concurrency of edges associated with a given node, measuring the probability that there are such edges overlapping in time, becomes by this IID assumption equivalent to the probability that any randomly-selected pair of links overlaps in time. We can thus select two randomly selected links with start times and durations denoted by $s_1, s_2$ and $d_1, d_2$. Without loss of generality, let $s_2 \geq s_1$. The probability that these two edges overlap in time is then simply the probability that the duration of the first edge is larger than the difference between starting times, $d_1 > s_2-s_1$. If we let $p(d)$ be the probability distribution function of durations and $I(s)$ be the probability distribution function of start times, the concurrency under these simplifying assumptions becomes
\begin{align}
    C &= 2\int_0^T I(s_1) \int_{s_1}^T I(s_2) \int_{s_2-s_1}^\infty p(d_1)\,dd_1\,ds_2\,ds_1\,.
\end{align}

\subsection{Simulated timings and temporal concurrency}
\label{subsec:exp_conc}

For the simulated temporal network data studied here, we further simplify the above expressions for temporal concurrency by assuming specific probability distribution functions for the start times and durations of edges, inherently nondimensionalizing the time scale of the two distributions so as to work easily within a one-parameter model for modifying temporal concurrency. In so doing, we emphasize that the temporal concurrency and the subsequent calculation of reachability depends only on the orderings of start and end times, not the total amounts of time involved in those overlaps. We take a uniform distribution for edge start times, $I(s)=1/T$, $s\in[0,T]$, and draw durations from an exponential distribution $p(d)=e^{-d}$. 

We emphasize that changing the decay rate of the exponential in $p(d)$ is unnecessary, since doing so is nondimensionally equivalent to a change in $T$ for calculating concurrency and reachability. That is, the range $T$ has inherently become a  nondimensional ratio of the underlying time scales of the distributions of start times to that of edge durations.

The cumulative distribution function of edge durations larger than some specified time, which we notate by $D(t)$, then simplifies to $D(t)=\int_{t}^{\infty}e^{-\tau}d\tau=e^{-t}$ and the concurrency of the temporal network under these timings can be rewritten as 
\begin{equation}
\label{eq:concurrency}
    C=\frac{2}{T^2}\int_{0}^{T}\int_{s_1}^{T}  e^{-(s_2-s_1)}\,ds_2\,ds_1 = \frac{2}{T^2}(T-1+e^{-T}).
\end{equation}

We note that $C \approx 2/T$ for $T \gg 1$. Meanwhile, taking the series expansion of the exponential, we obtain that the temporal concurrency for $T \ll 1$ approaches the value $1$ like $C \approx 1 - {T}/{3}$. For comparison, \eqref{eq:concurrency} gives $C(1) \doteq 0.736$.

Importantly, our structural cohesion definition depends only on the topology of the aggregated time-independent network, ignoring start and end times and including all edges that ever exist during the time period. In contrast, the temporal concurrency depends only on the distributions of start times and edge durations, independent of the network topology.

\subsection{Construction of synthetic temporal networks}
\label{subsec:construc_network}

To explore the effect of temporal concurrency and its interplay with structural cohesion, we will examine the reachability with a model approximation based on the assumption that the networks are locally tree-like. We thus start by confirming the analysis on balanced and unbalanced tree networks, which have only one node independent path for each node pair.
In the balanced tree networks, each node has $m$ successors except the leaves that are at distance $h$ from the root. 

We generate unbalanced tree networks by rewiring the balanced trees, ensuring that they maintain the same numbers of nodes and edges.
In our rewiring, we choose a random edge $(i,j)$ from the set of edges $\{\mathcal{E}\}$.
Removing this edge separates the network into two components: one includes $i$ and the other includes $j$. We then choose a random node $v$ from the component containing node $j$, and connect $i$ to $v$.
In so doing, we ensure at each step that we maintain a tree structure without cycles.
We continue this process $\phi|\mathcal{E}|$ times, where the rewiring fraction used in our work here is $\phi=0.1$; that is, we rewired 10\% of the links.

We generated connected components from Erd\H{o}s-R\'enyi (ER) networks using the {\tt gnp\_random\_graph function} in the NetworkX python package~\cite{team2014networkx}, 
Generating 100 ER networks initially from $N=120$ nodes and connection probability $p=0.017$ yielded largest connected components of size $\langle N \rangle=93.92\pm8.6$ and average degree $\langle k \rangle \approx 2$. For $p=0.025$, with the same initial size, 
the largest connected components have an average size of $\langle N \rangle=113\pm3.1$ and $\langle k \rangle \approx 3$.

We further compare these results with randomly generated graphs with exponential degree distributions, as described in the Appendix. In particular, we observe that the average structural cohesion for an exponential degree distribution graph is typically smaller than for an ER network with the same mean degree. We then subsequently rewire the ER and exponential degree networks to a desired, matched structural cohesion, to clarify the comparison being considered (see the Appendix).

To connect our results to the previous work of~\cite{Moody2016}, we used the same sampled collaboration networks studied there, which were extracted by four-step random walks from collaboration networks~\cite{Moody2004ASR}.
In particular, we consider the same four examples that were highlighted in Fig.~3 and Appendix~2 of~\cite{Moody2016}, having similar sizes to one another but different structural cohesion. These selected networks capture low average numbers of partners and skewed degree distributions, both of which are typical in sexual contact patterns, making them useful for testing the impact of temporal concurrency in the context of a spreading infection~\cite{Moody2016}.

For each of our four different classes of aggregated network structure (trees, ER, exponential degree distributions, and empirical examples), we randomly generate the temporal information for each edge (i.e., start times and durations). We note that we treat all of our synthetic networks as single-interval temporal networks, where each edge is present for the entirety of the duration after its start time, as drawn from the selected distributions. Given the start time, $s$, and duration, $d$, of an edge, its end time, is of course $\epsilon = s+d$. As described above, we consider start times drawn from a uniform distribution $s \in [0,T]$, with durations following an exponential distribution $p(d)=e^{-d}$. As such, $T$ is effectively a dimensionless time, which we vary in the range $T \in [0,20]$.

\subsection{Numerical reachability  by using accessibility matrix}
\label{subsec:numeric_reach}

Given the specific temporal contact information of every link in the temporal contact network, we directly evaluate the average reachability as the density of the accessibility graph. Direct contacts like $\rm(A,B)$ in Fig.~\ref{fig:schem}(a) immediately carry over into the accessibility graph, along with additional ordered pairs like $\rm(A,C)$ and $\rm(D,C)$ in Fig.~\ref{fig:schem}(c). For example, an infection starting from D at $t=1$ can reach C either by directly infecting B, or by infecting A who then infects B, and then by B infecting C during their (later in time) contact. However, an infection seeded at C cannot reach A or D because of the absence of temporally consistent paths, since the $\rm(B,C)$ link does not appear until after all of the other edges have ended. 

The role of concurrency as a potential enhancer of reachability is immediately apparent in this small toy example if we vary the start time of the $\rm(B,C)$ edge: if that start time were before $t=4$, then the ordered pair $\rm(C,D)$ would also be in the accessibility graph, so an infection seeded at $C$ can spread further than with the timings indicated in the figure. Similarly, if that start time were before $t=3$, the ordered pair $\rm(C,A)$ would also be accessible.

We describe the unweighted accessibility graph through its adjacency matrix $\+{R}$ with elements ${R}_{ij}=1$ when there is a temporally consistent path from node $i$ to node $j$, otherwise ${R}_{ij}=0$, as shown in Fig.~\ref{fig:schem}(e).
To quantify an average accessibility across the whole network, we calculate the density $R$ of the accessibility graph (that is, the density of the off-diagonal elements of the accessibility matrix)
\begin{equation}
\label{eq:R}
    R = \dfrac{1}{N(N-1)}\sum_{\substack{i\neq j}}R_{ij}\,,
\end{equation}
and we call this quantity $R$ the reachability of the temporal network.

\begin{figure}
\centering
\includegraphics[width=0.9\linewidth]{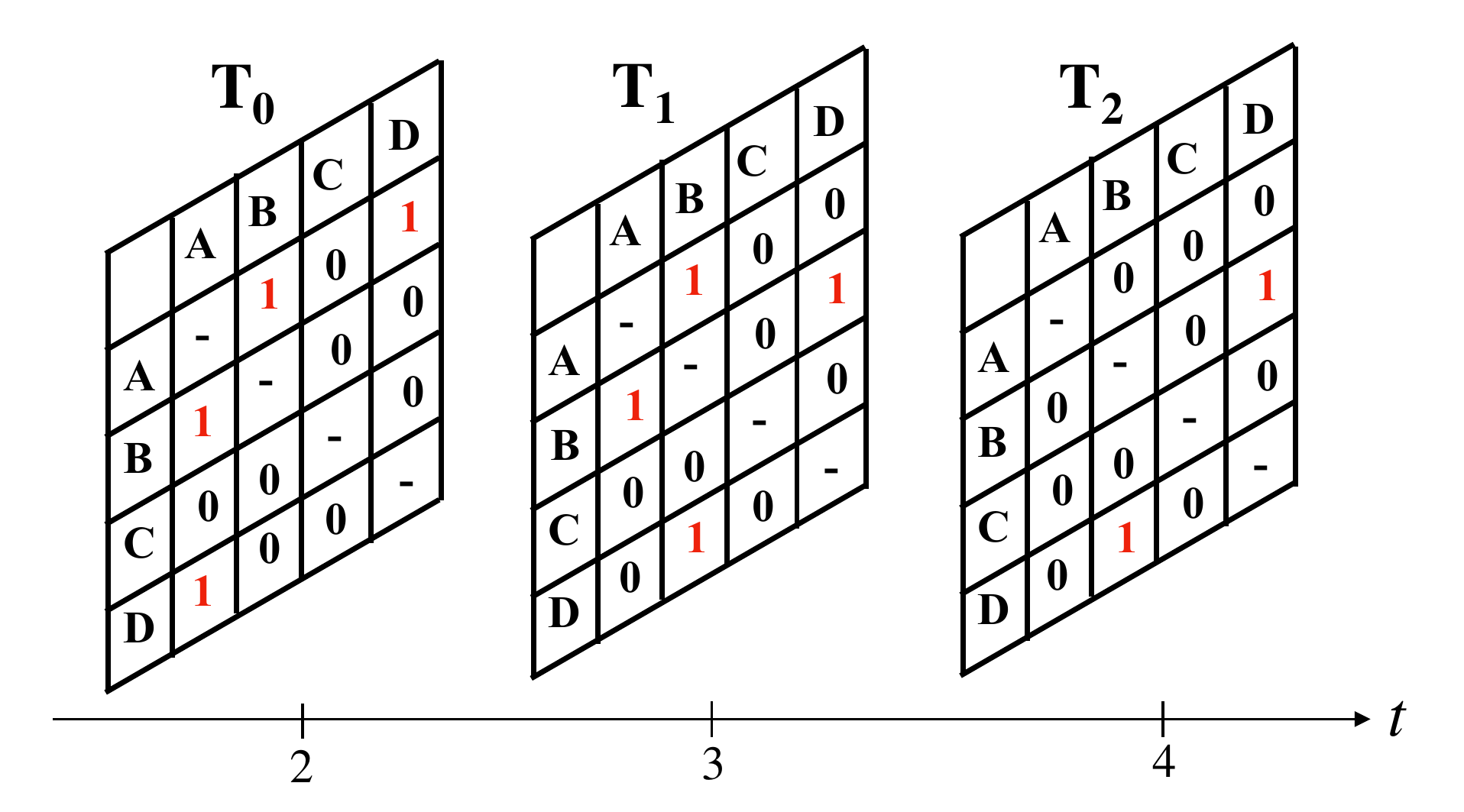}
\caption{Temporal layered matrices as observed immediately before the end time of a link. $\+T_{0}$ is the adjacency matrix immediately before the disappearance of the earliest ending link (A,D) from Fig.~\ref{fig:schem}. Similarly, $\+T_{1}$ is the adjacency matrix immediately before the end of the next ending link (A,B). That is, $\+T_{w}$ includes all links in Fig.~\ref{fig:schem} that overlap with the disappearance of the $w$th link (counting from $0$, $\+T_3$ not shown). If multiple edges end at the same time, the accessibility calculation proceeds equivalently either with a separate but identical matrix for each edge ending at that time, or a single matrix corresponding to that time.}
\label{fig:layer}
\end{figure}

To numerically evaluate the reachability from temporal network data, we represent the essential temporal information into layers of contacts corresponding to edge end times that have been sorted in ascending order, as depicted in Fig.~\ref{fig:layer}. The process of generating the temporal layers is as follows:
\begin{enumerate}
\item Sort edges in $\{\mathcal{E}\}$ by end time $\epsilon_w$ in ascending order, where $\epsilon_w$ is the end time of edge $l_w$, $w \in [0,E-1]$ is the sorted index of edges, and $E$ is the total number of edges, $E = |\mathcal{E}|$. For example, $l_0$ is the edge with the earliest end time, whereas $l_{E-1}$ is the last edge to end. (Breaking ties between identical end times is unimportant for calculating reachability, except insofar as it can be used to speed up the calculation by indexing the smaller number of distinct end times, under an appropriate change of notation.)

\item Construct the $w$th temporal layer matrix $\+T_w$ by including edge $l_w$ and all other edges $l_{w'}$ with $w'>w$ that are present just before the end time $\epsilon_w$. That is, $\+T_w$ includes $l_w$ and all $l_{w'}$ satisfying both $s_{w'} < \epsilon_w$ and $\epsilon_{w'}\geq\epsilon_{w}$.

\item By repeating step 2, the full set of temporal layer matrices $\+{T}_0,\+T_1, \dots, \+T_{E-2}, \+T_{E-1}$ may be prepared. 

\item Multiply the matrix exponentials of each\ temporal matrix $\+T_w$ to obtain $\displaystyle {\+R} = \prod_{w=0}^{E-1} e^{\+T_w}$. 
\item Binarize ${\+R}$: For all $R_{ij} >0$ values, set $R_{ij}=1$. 
\item Evaluate the average reachability ${R}$ by Eq.~\ref{eq:R}.
\end{enumerate}

For example, in Fig.~\ref{fig:schem}, the earliest ending edge $l_0 = (\rm{A,D})$ ends at time $\epsilon_0 = 2$. The edge $\rm{(A,B)}$ is the only other edge present in the temporal layer $\+T_0$ in Fig.~\ref{fig:layer}, satisfying the step 2 conditions above. One can similarly determine the adjacency matrices corresponding to the end time of each edge, and multiply the matrix exponentials to evaluate the accessibility matrix as described in step 4. Once we binarize the accessibility matrix $\+R$, the reachability $R$ is obtained by averaging the off-diagonal elements of $\+R$. In the example in Figs.~\ref{fig:schem} and \ref{fig:layer}, the reachability is ${R}=10/12\doteq 0.83$. 

The matrix exponentials in step 4 above provide a simply-expressed formula indicating the connected components within each individual temporal layer. As such, multiplying the matrix exponentials for any set of consecutive temporal layers yields (after binarizing) the reachable network associated with that combination of layers. But in practice for larger temporal networks, it is significantly more efficient computationally to instead directly calculate the connected components of $\+T_w$ and replace the matrix exponential with the binary indicator matrix whose elements specify whether the corresponding pair of nodes are together in the same component at that time. Similarly, to save memory overhead, steps 3 and 4 can be trivially combined to consider only one temporal layer at a time. For even larger networks whose adjacency matrices must be represented as sparse matrices to fit in memory, the corresponding accessibility graph could instead be constructed one row at a time, updating the running average of the density $R$ to calculate the overall reachability.


\section{Approximate model for reachability}
\label{sec:model_reach}

We seek to approximate the reachability in terms of some minimal temporal and structural information necessary to accurately describe the essential relationships. Using simulated timings on random graphs, we immediately observe that the overall density and level of cohesion are insufficient for describing the needed structural effects. Specifically, we consider simulated timings on random networks with exponential degree distributions and ER graphs that have been rewired to target specific cohesion values as described in the Appendix. The results in Figure~\ref{fig:erexp_control_cohesion} show reachability versus concurrency for rewired ER and exponential degree distribution graphs both with $\langle k\rangle=3$ and $\langle\kappa\rangle=1.7$, demonstrating clear differences. As such, we desire to more accurately approximate the reachability versus concurrency relationship using more structural information. Motivated by Fig.~\ref{fig:erexp_control_cohesion} and modeling successes for other network problems (see, e.g., \cite{Melnik2011PRE,Gleeson2012,Gleeson2013PRX} and citations therein), we consider approximations developed in terms of the underlying degree distribution.

We develop a heterogeneous mean-field model for reachability specified in terms of the degree distribution $p(k)$, temporal concurrency, and structural cohesion. In so doing, we implicitly assume that the underlying aggregated network is sufficiently locally tree-like (see, e.g.,~\cite{Melnik2011PRE}). Assuming that the essential network structure is (at least largely) dictated by the degree distribution $p(k)$, we proceed to develop models for the effects of the temporal concurrency of edges. 

As above, let $I(t)$ be the probability of edge starting times, and $D(t)$ be the probability of an edge duration larger than $t$. We seek the probability $P(s,\ell)$ that a given chain of $\ell$ edges is temporally consistent given that the first edge has start time $s$. By definition, $P(s,1)=1$, since any edge considered in isolation is temporally consistent with itself. 
Assuming all edge start times and durations are independent and identically distributed, the recursive equation for $P(s,\ell)$ is developed by considering whether the start time of the next edge in the chain is before or after $s$: 
\begin{equation}
\label{eq:psn}
    P(s,\ell)=\int_{0}^{s}\!I(t)D(s-t)P(s,\ell-1)\,dt +\int_{s}^{T}\!I(t)P(t,\ell-1)\,dt\,.
\end{equation}
The first integral accounts for the possibility that the next edge in the chain has a start time before the first edge, distributed according to $I(t)$, but duration long enough to be concurrent with the first edge, with probability $D(s-t)$. Importantly, we then assess that this next edge is the first edge in a chain of $(\ell-1)$ edges that are temporally consistent with probability $P(s,\ell-1)$, as opposed to $P(t,\ell-1)$, to account for the requirement that all of the edges on the remaining chain still need to be concurrent or appear after the original edge start time $s$. In contrast, the second integral directly measures the contribution from the next edge starting at time $t>s$, after the start time of the first edge, along with the probability that an edge starting at time $t$ is part of a chain of $(\ell-1)$ edges that are temporally consistent. The second integral here ends at time $T$ since by our definition $I(t)=0$ for $t>T$.

Given $P(s,\ell)$, we can then determine the probability $W_\ell(T)$ that a randomly selected chain of edges of path length $\ell$ is temporally consistent, by averaging over the $I(s)$ distribution for the start time of the first edge:
\begin{equation}
\label{eq:Wn}
    W_\ell(T)=\int_{0}^{T}I(s)P(s,\ell)\,ds\,,
\end{equation}
where we have here explicitly noted the remaining dependence on $T$ for temporal consistency along path length $\ell$. We note in particular that the initialization $P(s,1)=1$ yields $W_1(T) = 1$ for all $T$; that is, a path of length $1$ is necessarily temporally consistent.

\subsection{Consideration of node-independent paths}
\label{subsec:Rbar}

Motivated by Moody and Benton \cite{Moody2016} --- specifically, inspired by their observations about the role of structural cohesion as measured by the average number of node-independent shortest paths --- the development of our model approximation proceeds by restricting attention to the node-independent shortest paths between a node pair $(i,j)$. This treatment, taking the assumption of locally tree-like structure to an extreme, allows us to treat the probabilities along each node-independent path independently. But in doing so, we recognize that we will undoubtedly fail to take into account all potential detours around temporally-inconsistent parts of these paths. Nevertheless, as we will see below, this approach appears to be relatively accurate for small enough structural cohesion and particularly so at low levels of concurrency, presumably because the probabilities $W_\ell(T)$ drop off quickly with increasing $\ell$ under such conditions.

As above, we continue to denote the number of node-independent shortest paths between nodes $i$ and $j$ by $\kappa(i,j)$. We index those paths by $q\in\{1,\cdots,\kappa(i,j)\}$ and identify the length of path $q$ by $\ell(q)$. We then seek the probability that path $q$ is temporally consistent, which we write as $W_{\ell(q)}(T)$. By the definition of reachability, an ordered node pair is accessible if there is at least one temporally-consistent path from the one node to the other; so we only need to exclude the case that there are no temporally-consistent paths between the nodes. Continuing to assume that we can reasonably consider only the node-independent paths, the probability $\rho(i,j)$ that at least one of these $\kappa(i,j)$ node-independent shortest paths under consideration between $i$ and $j$ is temporally consistent follows simply by independence:
\begin{equation}
\label{eq:rho}
\rho(i,j) = 1 - \prod_{q=1}^{\kappa(i,j)}\left[1-W_{\ell(q)}(T)\right]\,.
\end{equation}

That is, given the computation~\cite{Newman2001} that separately identifies the number and length of node-independent shortest paths for each node pair $(i,j)$ in the aggregated network, equation \eqref{eq:rho} gives us the probability of accessibility between the pair, as restricted along these node-independent paths. In other words, the corresponding $R(i,j)$ element of the accessibility matrix becomes 1 with probability $\rho(i,j)$. The expected density of the accessibility graph (the off-diagonal parts of the accessibility matrix) under our approximation thus becomes
\begin{equation}
\label{eq:barR}
\bar{R} = \frac{1}{N(N-1)}\sum_{i\neq j} \rho(i,j)\,.
\end{equation}

By our construction, $\rho(i,j)=\rho(j,i)$, though this does not require corresponding similarity in the elements of $\+R$. 
We also note that we would ideally consider edge-independent paths, which are by definition at least as numerous as the node-independent ones. But given the observed relationship between structural cohesion and the degrees of node pairs in our random graph results in Fig.~\ref{fig:S1_cohesion} in the Appendix, we expect that the typical numbers of edge-independent shortest paths should not on average be much greater than the node-independent ones in these random cases.

\subsection{Modeling in terms of the distribution of path lengths}
\label{subsec:Rtilde}

The above calculation of $\bar{R}$ requires detailed knowledge of the number and lengths of the node-independent paths between each $(i,j)$ pair. That is, for all intents and purposes we need the entire structure of the aggregated network upon which to calculate these quantities. However in many network survey conditions, the available information is much more tightly constrained. It can be particularly beneficial under such settings to model outcomes at the level of heterogeneous mean-field theories that use only the degree distribution of the network (see, e.g., \cite{Satorras2001PRL,Gleeson2013PRX}). Since such models are typically derived from ``locally tree-like" assumptions (see, e.g.,~\cite{Melnik2011PRE}), we find it reasonable to consider how we might similarly extend our tree-like assumptions above.

Given a distribution of path lengths, $p(\ell)$, to be considered as independent candidate paths between a randomly selected pair of nodes, the joint probability that a selected path has length $\ell$ and is temporally consistent is given by $p(\ell)W_\ell(T)$. Summing over possible path lengths, we compute the probability $\hat\rho$ that a path from this set is temporally consistent: 
\begin{equation}
\label{eq:q}
    \hat\rho = \sum_{\ell=1}^{L}p(\ell)W_\ell(T)\,, 
\end{equation}
where $L$ is the largest path length in the distribution $p(\ell)$. Then the probability that at least one of $\kappa(i,j)$ independent paths between nodes $i$ and $j$ is temporally consistent simplifies to
\begin{equation}
\label{eq:minkk}
  \tilde\rho(i,j) = 1-(1-\hat\rho)^{\kappa(i,j)}\,.
\end{equation}
Notably, using the probability $\hat\rho$ is this way decouples the considered probabilities along each path from all other possible properties of importance of nodes $i$ and $j$ (e.g., their degrees). And, again, in making this calculation we have made the (rather strong) assumption that we considered only independent paths.

To model $\kappa(i,j)$, we note that while an exact analytical measure of the structural cohesion appears to be prohibitively difficult, a trivial application of Menger's theorem~\cite{Menger1927} requires the maximum number of node-independent paths between nodes $i$ and $j$ to be bounded by the minimum degree of the pair, $\min(k_i,k_j)$, where $k_i$ and $k_j$ indicate the degrees of the two nodes.
We observe that this upper bound yields a good approximation for the average cohesion $\langle\kappa\rangle$ in our random graphs, as observed in Fig.~\ref{fig:S1_cohesion}. That said, we note by way of contrast that the four empirical networks from~\cite{Moody2016} that we study have much lower cohesions (1.61, 1.34, 1.07 and 1.06) than bounded by this relationship to node degrees (3.18, 3.39, 2.10 and 2.01, respectively). Moreover, in a true tree we require $\kappa(i,j)=1$ for all node pairs by definition.

By assuming $\kappa(i,j)\approx \min(k_i,k_j)$ and substituting the approximation into Eq.~\ref{eq:minkk}, $\tilde\rho(i,j)$ depends only on the node degrees $k_i$ and $k_j$, along with the path length distribution $p(\ell)$ under consideration. The resulting approximation of reachability, denoted $\tilde{R}$ to distinguish it from the $\bar{R}$ calculation of the previous subsection, then becomes
\begin{equation}
\label{eq:tildeR}
    \tilde{R} = \sum_{k_i,k_j}p(k_i)p(k_j)\tilde\rho(i,j)\,,
\end{equation}
where $p(k)$ is the degree distribution and we have not bothered to correct the $O(1/N)$  contribution in $\tilde{R}$ corresponding to pairing a node with itself.

We again emphasize that we have assumed the structural and temporal details of our temporal networks are independent of one another. Therefore, for instance, there are no correlations between node degrees and all of the temporal details absorbed into the $W_\ell(T)$ terms. The only remaining structural contributions in the $\tilde{R}$ approximation are from (1) the empirically observed $p(k)$ degree distribution, (2) the selected model for $\kappa(i,j)$ as discussed above, and (3) the selected distribution of path lengths $p(\ell)$ to obtain $\hat\rho$ in Eq.~\ref{eq:q}. 

In theory, one could continue by way of approximating $p(\ell)$ in terms of the degree distributions~\cite{Katzav2015EPL, Mor2016PRE}. In particular, we note that in going in this direction one is more likely to be able to employ some model for the distribution of geodesic shortest path lengths, as opposed to that for node-independent shortest paths. Similarly, if the path length distribution is to be sampled by some manner, it may be more likely to get a reasonable sample of the geodesic paths versus the node-independent ones. To explore the effect of potentially using the geodesic shortest path length distribution instead of the node-independent path length distribution, we below consider both possibilities by direct use of the empirically observed path length distributions in each network, using $\tilde{R}_s$ to represent the approximation obtained using the shortest path length distribution and $\tilde{R}_{n}$ for the model using the node-independent path length distribution.  

\section{Results}
\label{sec:results}
We numerically examine the relationship between temporal concurrency and reachability on different families of networks, comparing with our model approximations. The reachability approximation $\bar{R}$ from Sec.~\ref{subsec:Rbar} uses the specific structural information of numbers and lengths of node-independent paths between each node pair. In contrast, the $\tilde{R}_s$ and $\tilde{R}_n$ approximations from Sec.~\ref{subsec:Rtilde} employ path length distributions over the whole network, using the distributions of shortest paths and of node-independent paths, respectively. We confirm the results for trees (balanced and unbalanced). We then test the calculation on Erd\H{o}s-R\'enyi networks at two different densities and on four empirical networks highlighted in \cite{Moody2016}.

\subsection{Reachability on synthetic tree networks}

We numerically evaluate the reachabilty and our approximation, varying the temporal concurrency $C$ on balanced and randomly unbalanced tree networks with two different sizes (specified by the number of offspring, $m$, and the depth, $h$, of the balanced tree): (i) $m=2$ and $h=6$, with $N=127$ nodes; and (ii) $m=3$ and $h=4$, with $N=121$ nodes.
The average degrees of these two types of trees are $\langle k \rangle \doteq 1.98$.
We numerically evaluate the reachability ${R}$ by the method in Sec.~\ref{subsec:numeric_reach} and compare it with the $\bar{R}$ (Eq.~\ref{eq:barR}) and $\tilde{R}$ (Eq.~\ref{eq:tildeR}) approximations. Since, by definition, a tree provides only a single node-independent path between a node pair, we accordingly set $\kappa(i,j)=1$ in Eqs.~\ref{eq:rho} and \ref{eq:minkk} in calculating $\bar{R}$ and $\tilde{R}$, respectively. Similarly, because the node-independent and geodesic shortest path distributions are thus identical, we note that $\tilde{R}_n=\tilde{R}_s$ on a tree.

\begin{figure}[ht!]
\centering
\includegraphics[width=0.98\linewidth]{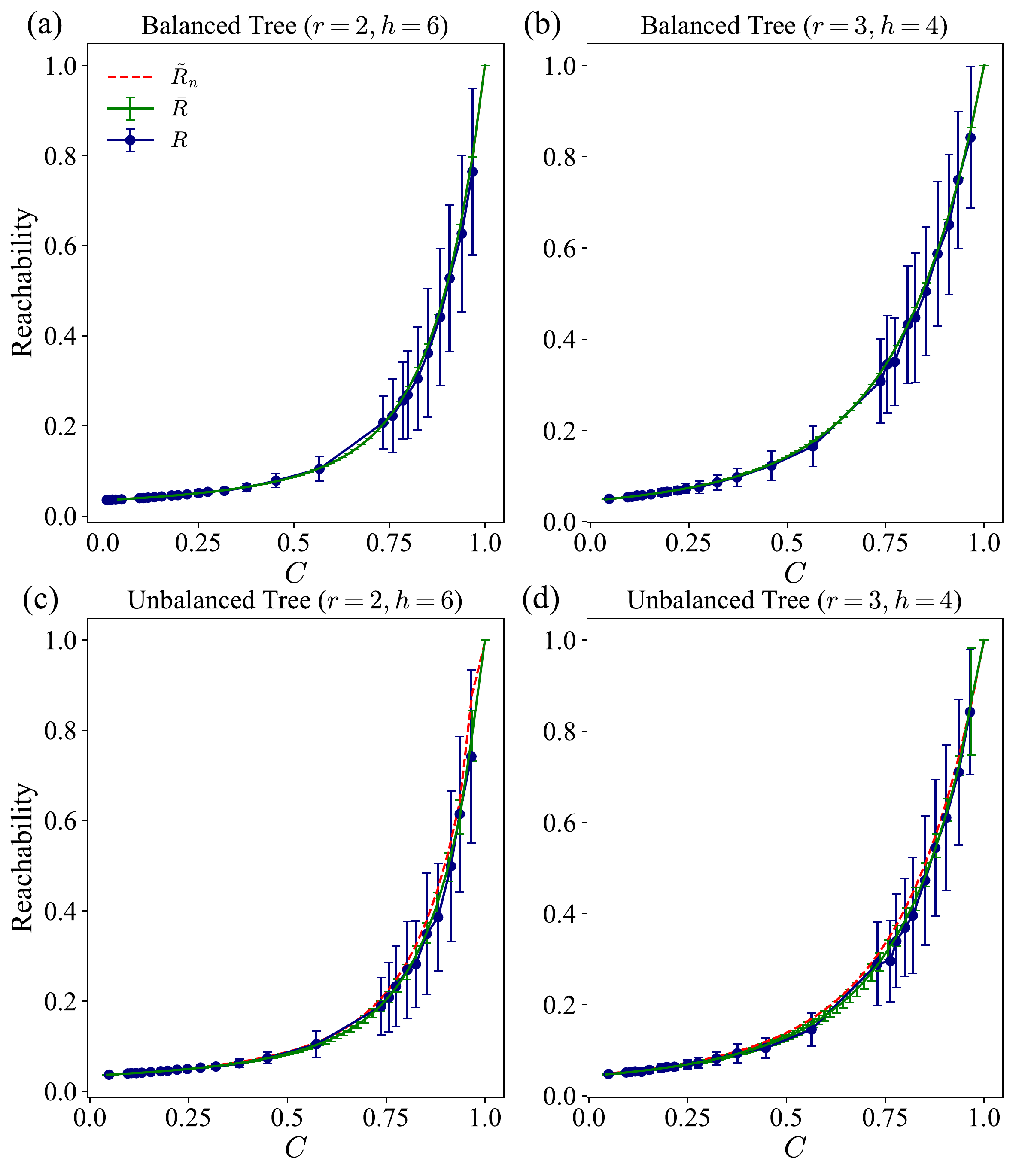}
\caption{Simulated and approximated reachability on balanced and unbalanced tree networks as a function of temporal concurrency $C$. Two different sizes of trees are tested: (a) and (c) are for $m=2$, $h=6$ ($N=127$) and (b) and (d) are for $m=3$, $h=4$ ($N=121$). Navy lines display the simulated reachability with $100$ synthetic balanced and unbalanced networks. Green dashed lines represent the approximation $\bar{R}$ relying on numbers and lengths of paths between each node pair. The dashed red line shows the approximated reachability with the node-independent path length distribution $\tilde{R}_n$ (which is equivalent to $\tilde{R}_s$ on a tree). Error bars indicate standard deviations over the different rewired unbalanced trees.} 
\label{fig:result_tree}
\end{figure}

In Fig.~\ref{fig:result_tree}, the approximations accurately describe the typical increase in reachability with increasing temporal concurrency for both the balanced and unbalanced trees. We specifically note that the concurrency values plotted here are the expected value given a specified time interval $T$. The results on the unbalanced trees include different network realizations as obtained by the rewiring described in Sec.~\ref{subsec:construc_network}. We observe a very slight gap between the approximations $\bar{R}$ and $\tilde{R}_n$ for the unbalanced trees, compared to that of the balanced trees, though the predictions are still well within the standard deviation of the data. We hypothesize that the greater heterogeneity in the path length distribution $p(\ell)$ in the unbalanced tree network may be a possible cause of this difference. The result confirms that the approximations accurately estimate the reachability for tree networks using only the path length and degree distributions $p(\ell)$ and $p(k)$, which is of course expected in this setting.

\subsection{Reachability on Erd\H{o}s-R\'enyi networks}
\label{subsec:reach_ER}

We next consider the reachability on sparse Erd\H{o}s-R\'enyi (ER) networks with $\langle k \rangle \approx 2$ and $3$, going in with the assumption that these random graphs will typically have sufficiently locally tree-like structure~\cite{newman2010networks}. 

\begin{figure}[ht!]
\centering
\includegraphics[width=0.98\linewidth]{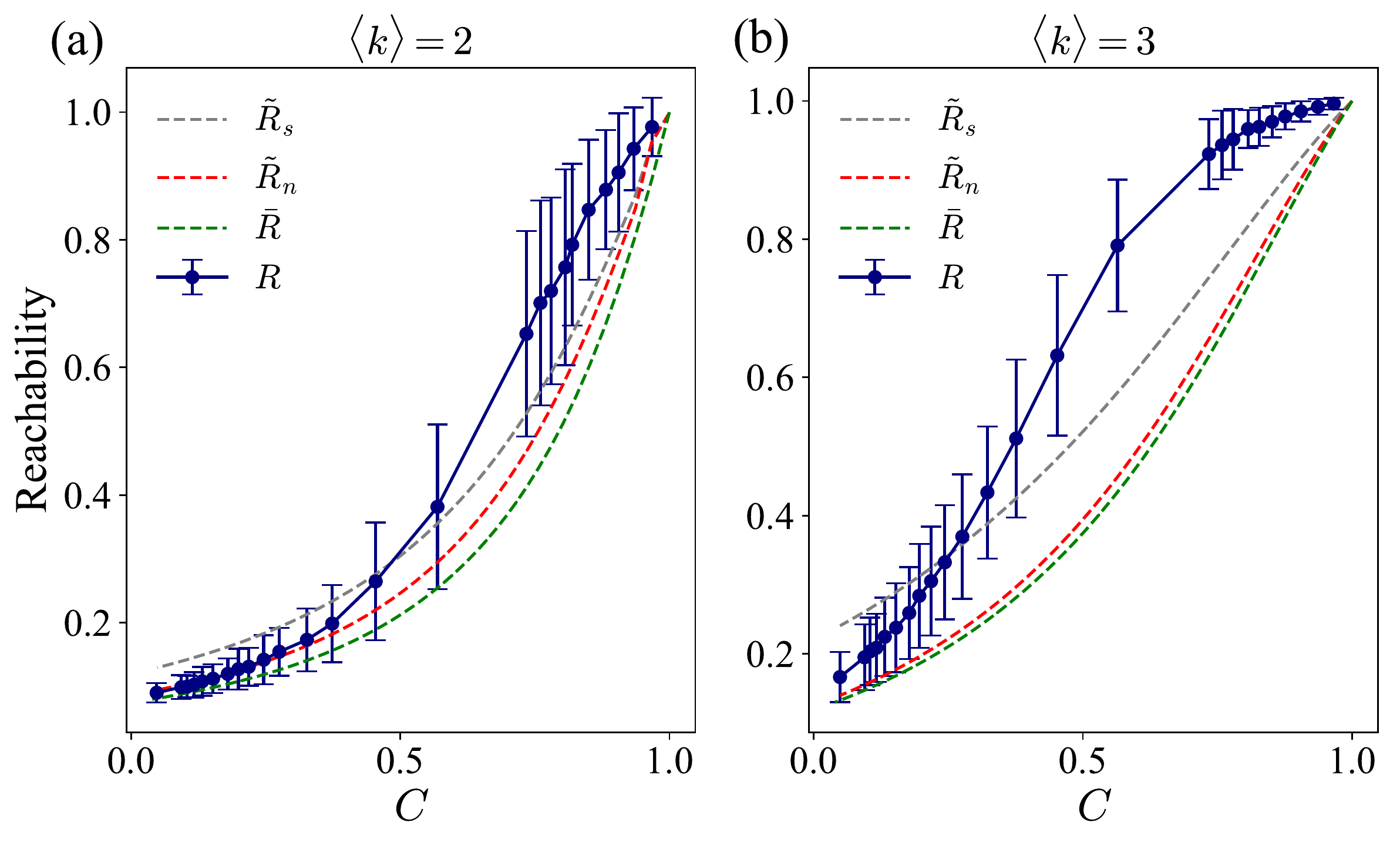}
\caption{Simulated and approximate reachability on sparse ER networks as a function of temporal concurrency $C$. Largest connected components for two different size networks are tested: (a) $\langle k \rangle \approx 2,  \langle N \rangle=94$, and (b) $ \langle k \rangle \approx 3, \langle N \rangle=113$. The solid line indicates the numerically measured reachability simulated over $100$ ER networks for each size. Error bars indicate standard deviations. Dashed lines indicate the different approximations for the reachability.}
\label{fig:result_ERtree}
\end{figure}

As shown in Fig.~\ref{fig:result_ERtree}(a), the approximation $\bar{R}$ that includes the specific path length information between node pairs in the network largely underestimates the reachability. If anything, we should not be surprised that $\bar{R}$ underestimates the true value of $R$ like this, since the calculation leading to $\bar{R}$ only considers reachability along node-independent paths. As such, the increased error made by $\bar{R}$ in increasing from $\langle k\rangle\approx 2$ to $3$ is expected, though the size of the resulting error emphasizes the apparent importance of available detours around these paths even for these small mean degrees. We note in particular that the $\bar{R}$ approximation is quite good at very low concurrency, $C$, where the node-independent shortest paths presumably have greater dominance because longer paths along detours become even more unlikely to maintain temporal consistency. However, the limiting behavior of the approximation in the $R\to 1$ approach as $C\to 1$ is clearly incorrect. We hypothesize that the behavior in this limit is possibly controlled by temporal inconsistency of key edge-to-edge transitions important along many paths, which is not an effect considered in the approximation.

We also include the approximations $\tilde{R}_n$ and $\tilde{R}_s$ in Fig.~\ref{fig:result_ERtree}. We note that $\tilde{R}_n$ is very similar to $\bar{R}$ here, indicating only modest change in the jump in the approximation obtained using full path length information for each node pair ($\bar{R}$) versus a single path length distribution $p(\ell)$ across all node-independent shortest paths ($\tilde{R}_n$). The additional gap between $\tilde{R}_n$ and $\tilde{R}_s$ is due to replacing the path length distribution empirically obtained over all node-independent shortest paths with the geodesic shortest path distribution, yielding shorter paths which are slightly more likely to be temporally consistent. Thus, $\tilde{R}_s$ slightly overestimates the reachability at very low temporal concurrency. 

\begin{figure}[ht!]
\centering
\includegraphics[width=0.98\linewidth]{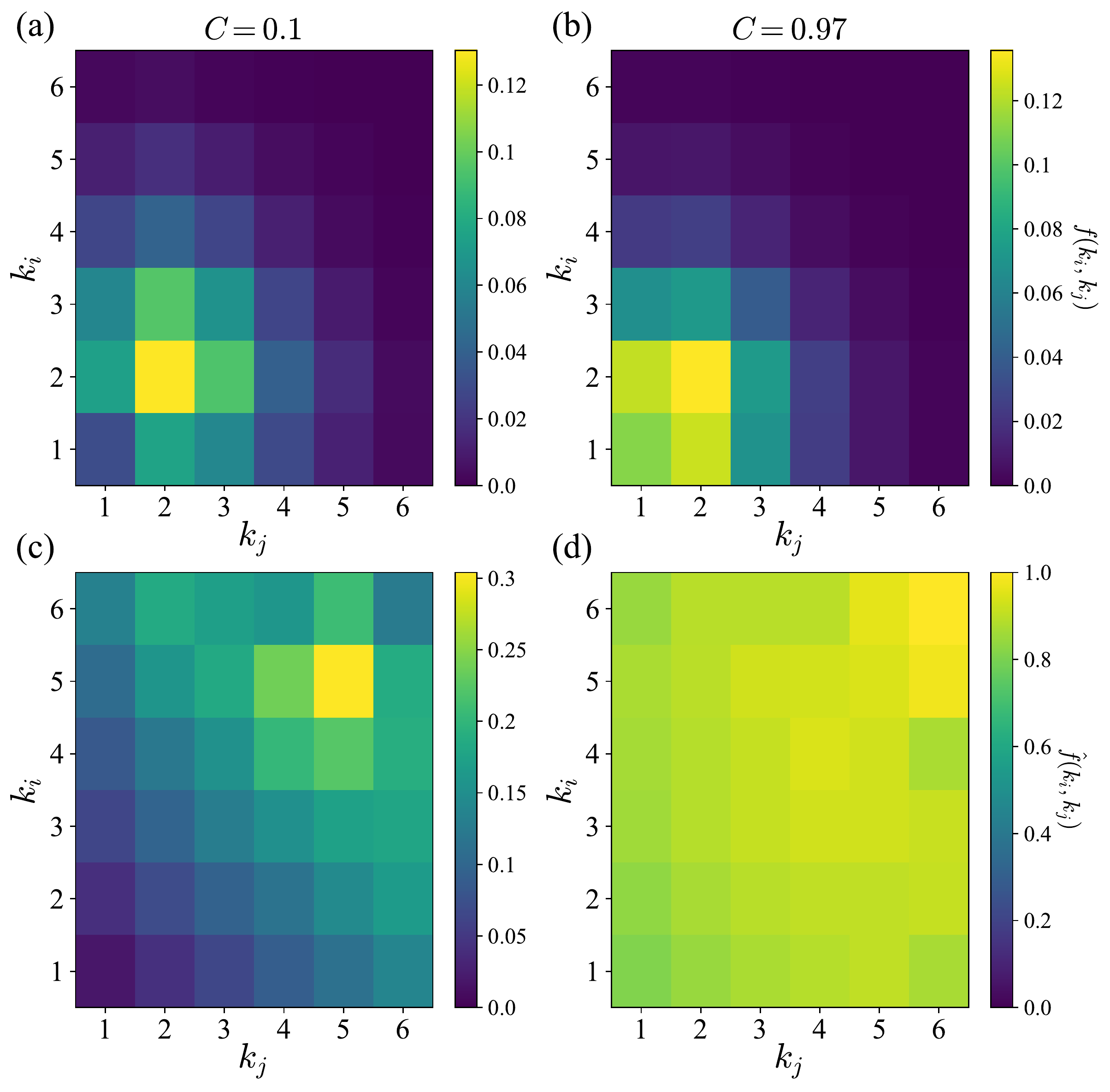}
\caption{Reachability frequency $f(k_i,k_j)$ and relative reachability frequency $\hat{f}(k_i,k_j)$ for ER networks with $\langle k \rangle \approx 2$. The numbers of reachable node pairs are counted from the accessibility matrix $\+R$ for the largest connected component of $100$ ER networks. The top panels, (a) and (b), display the reachability frequency $f(k_i,k_j)$, the fraction of reachable node pairs that have the indicated degrees. The bottom panels, (c) and (d), show the relative connectivity frequency $\hat{f}(k_i,k_j)$, the fraction of node pairs with the indicated degrees that are reachable. The temporal concurrency for the two left panels (a) and (c) is $C\doteq 0.1$. For the two right panels (b) and (d), $C\doteq 0.97$.}
\label{fig:result_ERtree_connect}
\end{figure}

To further understand the limitations of our approximations, we explored the reachability frequency of node pairs according to their degrees in ER networks with $\langle k \rangle \approx 2$. We directly measure how many node pairs with given degrees are reachable out of the total number of reachable node pairs:
\begin{equation}
\label{eq:regular_connnectivity}
    f(k,k')=\dfrac{\sum_{i\in\Lambda_k, j\in\Lambda_{k'},i\neq j} R(i,j)}{\sum_{i\neq j}R(i,j)}\,,
\end{equation}
where $\Lambda_k$ and $\Lambda_{k'}$ represent the sets of nodes having degree $k$ and $k'$, respectively. We measured $f$ across $100$ ER networks with $\langle k \rangle \approx 2$ for low [$C\doteq 0.1$ in Fig.~\ref{fig:result_ERtree_connect}(a)] and high temporal concurrency [$C\doteq 0.97$ in Fig.~\ref{fig:result_ERtree_connect}(b)]. 
Perhaps remarkably, we observe an only very small shift in $f$ between these two panels in the Figure, but the shift in the distribution that is apparent indicates that a larger fraction of the reachable pairs for high concurrency involve the degree-one nodes. That this should be the case makes intuitive sense in that the reachability of the degree-one nodes should be more suppressed at low concurrency than that for higher-degree nodes.

Noting that most node pairs are reachable at $C\doteq 0.97$, we make this observation more explicit by also computing the relative reachability frequency $\hat{f}(k_i,k_j)$ between degree pairs, defined as
\begin{equation}
\label{eq:relative_connnectivity}
    \hat{f}(k,k')=\dfrac{\sum_{i\in\Lambda_k,j\in\Lambda_{k'},i\neq j} R(i,j)}{\sum_{i\in\Lambda_k,j\in\Lambda_{k'},i\neq j}H(i,j)},
\end{equation}
where $\+H$ is the reachable matrix of the corresponding static network obtained by aggregating the temporal contacts. Because we only consider largest connected components in our numerical experiments, $H(i,j)=1$ and the sum in the denominator merely counts the number of such pairs given the selected degrees.
Fig.~\ref{fig:result_ERtree_connect}(c) and (d) shows the relative reachability frequency for low ($C\doteq 0.1$) and high ($C\doteq 0.97$) temporal concurrency, respectively. In particular, we confirm in panel (d) that almost all pairs are reachable, with $\hat{f} \approx 1$ for all degree values. In contrast, for $C\doteq 0.1$, the low-low-degree node pairs are much less likely to be reachable, as seen in Fig.~\ref{fig:result_ERtree_connect}(c)). Meanwhile, even at low concurrency, we see that high-high-degree node pairs are already quite likely to be reachable, with nearly $30\%$ of $(k_i,k_j)=(5,5)$ node pairs being reachable in this setting. We note in looking at Fig.~\ref{fig:result_ERtree_connect}(c)) that there are very few degree-$6$ nodes in these networks, so the apparent dropoff in $\hat{f}$ for these cases is due to averaging over a small number of such cases.

\begin{figure}[ht!]
\centering
\includegraphics[width=0.8\linewidth]{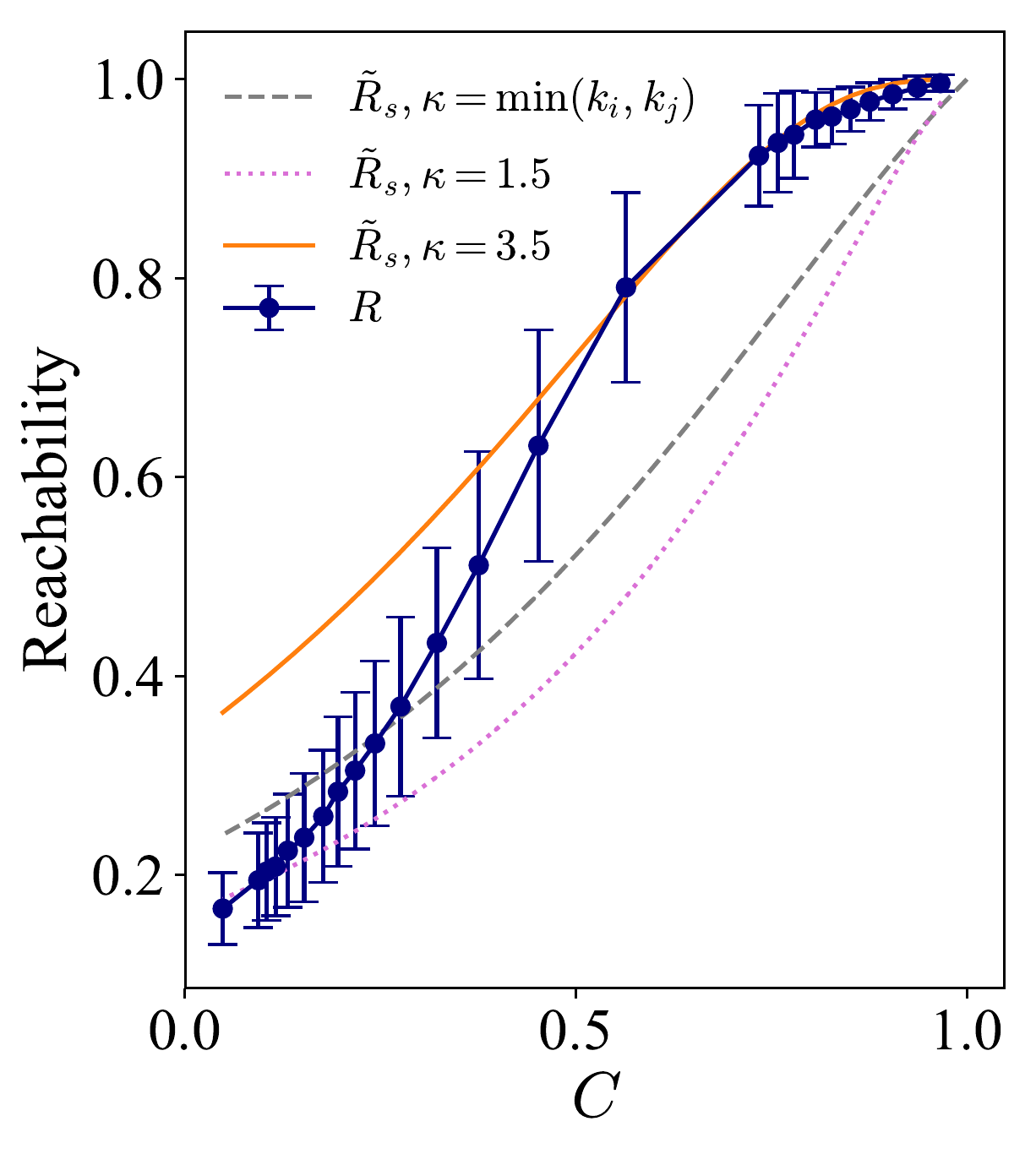}
\caption{Fit of $\kappa$ for the reachability of ER network largest connected components with $\langle k \rangle = 3$. The circle with error bar and grey dashed line represent the numerical reachability and $\tilde{R}_s$ approximation with the shortest path length distribution, respectively. The dotted line shows $\tilde{R}_s$ recalculated with constant $\kappa(i,j)=1.5$, while the solid curve uses $\kappa(i,j)=3.5$ as the exponent in Eq.~\ref{eq:minkk} for all node pairs.} 
\label{fig:result_ER_fit}
\end{figure}

The increasing errors in our model predictions at higher concurrency are directly because of the increasing importance of the neglected detours around the node-independent shortest paths. Recalling the explicit role of the number $\kappa(i,j)$ of such paths between nodes $i$ and $j$ in our approximations, we ask whether the relationship between reachability and concurrency observed numerically might be captured by assuming some other effective values for $\kappa$. In Fig.~\ref{fig:result_ER_fit}, we continue to consider reachability on the $\langle k\rangle\approx 3$ ER networks. Focusing for this figure only on $\tilde{R}_s$ approximations built from $p(k)$ degree distributions and $p(\ell)$ distributions of geodesic shortest paths, we reproduce here our regular $\tilde{R}_s$ approximation using $\kappa(i,j)=\min(k_i,k_j)$ from Fig.~\ref{fig:result_ERtree}(b). This approximation overestimates the reachability at low concurrency because the $p(\ell)$ distribution of geodesic shortest paths are shorter on average than the full set of node-independent shortest paths (the latter used in our $\tilde{R}_n$ approximations). As seen in Fig.~\ref{fig:result_ER_fit}, this overestimate at low concurrency can also be at least partially corrected for by decreasing the effective cohesion used in the approximation formulae to $\kappa=1.5$. (For comparison, the average structural cohesion of the underlying ER networks is $\langle\kappa\rangle\doteq 2.08$.) Of perhaps greater interest, we see in Fig.~\ref{fig:result_ER_fit} that the underestimated reachability at large $C$ appears to be corrected for at this level of modeling by choosing an effective cohesion value of $\kappa=3.5$, yielding a good approximation over the range $0.5\lesssim C\leq 1$. We believe that identifying such effective cohesion values as modeled from other network features (as opposed to curve fitting here) may be an interesting direction for future work, as a means of extending the range of validity of our tree-based approximations.

\begin{figure*}[ht!]
\centering
\includegraphics[width=0.95\linewidth]{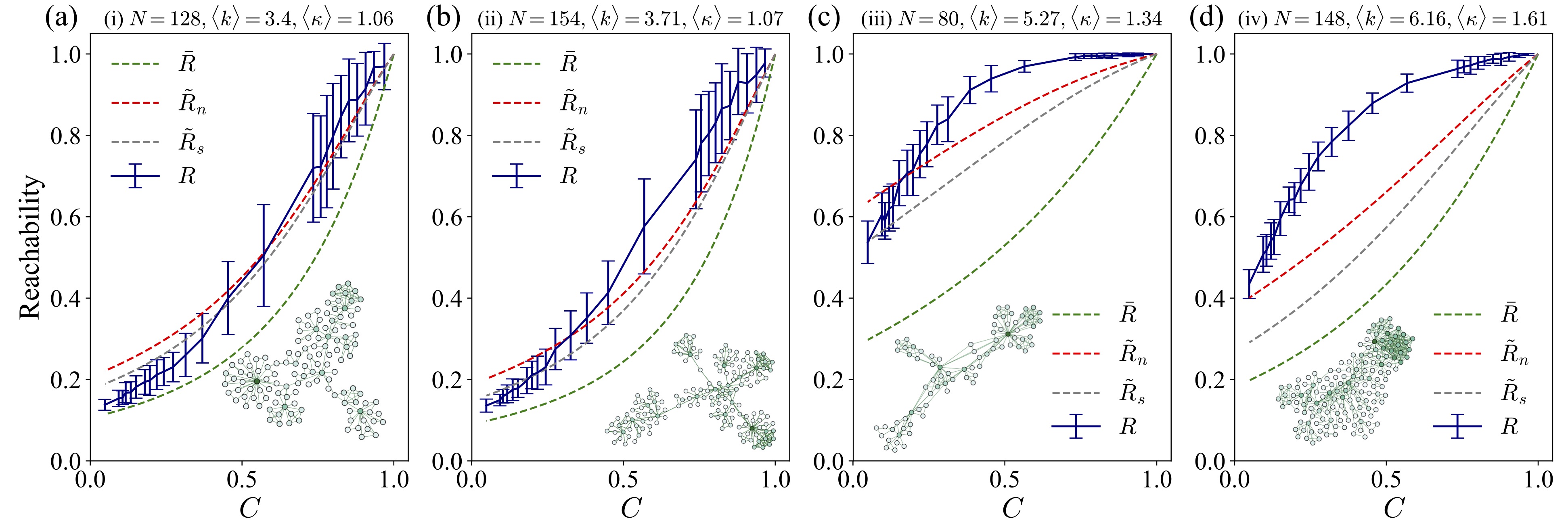}
\caption{Simulated and approximate reachability as a function of the temporal concurrency for empirical networks with different levels of structural cohesion: (a) sample network (i), $\langle \kappa \rangle\doteq 1.06$, $N=128$; (b) sample network (ii), $\langle \kappa \rangle \doteq 1.07, N=154$; (c) sample network (iii) $\langle \kappa \rangle\doteq 1.34$, $N=80$; and (d) sample network (iv), $\langle \kappa \rangle=1.61$, $N=148$. Means of simulated results $R$ are indicated by solid lines with error bars indicating standard deviations over $50$ different simulated timings. The green dashed line indicates the approximation $\bar{R}$ obtained from measured numbers and lengths of node-independent shortest paths between each node pair, as in Eq.~\ref{eq:barR}. The other dashed lines indicate the approximations $\tilde{R}_s$ and $\tilde{R}_n$ using path length distributions, as in Eq.~\ref{eq:tildeR}. Each network sample is visualized in the corresponding panel, with node shade indicating degree.} 
\label{fig:result_Empi_reach}
\end{figure*}

\subsection{Reachability on empirical networks}

We examined reachability versus concurrency on four sampled empirical networks that were highlighted in the previous work of \cite{Moody2016}.
Example networks (i) and (ii) have low structural cohesion --- $\langle \kappa \rangle \doteq 1.06$ and $\langle \kappa \rangle \doteq 1.07$, respectively --- while example networks (iii) and (iv) have relatively higher structural cohesion --- $\langle \kappa \rangle\doteq 1.34$ and $\langle \kappa \rangle \doteq 1.61$, respectively. 

When the network structure is tree-like in the sense of cohesion $\langle\kappa\rangle$ being near $1$, all three of our model approximations plotted in Fig.~\ref{fig:result_Empi_reach} appear to be in relatively good agreement with the numerically calculated reachability. In accord with our other results above, we see that our $\bar{R}$ approximation reasonably captures the low-concurrency limiting behavior in Fig.~\ref{fig:result_Empi_reach}(a,b), and while it necessarily underestimates the level of reachability throughout, the deviation from the true reachability curves at low structural cohesion (panels a and b of the Figure) are not as large as at higher cohesion (panels c and d). Moreover, we see that much of this underestimate is effectively corrected in this case by the other modeling steps introduced by the $\tilde{R}_n$ and $\tilde{R}_s$ approximations, again particularly so at lower values of cohesion. 

We also note here that the $\tilde{R}_n$ approximation overestimates the reachability in the low concurrency regime in panels a and b, unlike the above-observed behavior for ER graphs. This occurs because the way we constructed the empirical distribution $p(\ell)$ of the node-independent shortest paths for this calculation here counted multiple short paths between nearby nodes. This counting yields on average shorter paths that then overestimate the reachability at small concurrency.

\section{Conclusion}
\label{sec:conclusion}
We investigated the overall level of reachability in temporal networks, considering the effects of temporal concurrency and its interplay with network structure, including structural cohesion. We developed a sequence of approximations for reachability based on strong (and potentially inaccurate) assumptions of locally tree-like networks. We then compared our approximations to numerical results for simulated edge timings on a variety of types of networks.
In networks that are tree-like in the sense of low structural cohesion, our approximation agrees well with the numerically computed reachability, particularly so for small concurrency. At larger structural cohesion and/or larger concurrency, the importance of having many possible non-independent paths is not captured by our existing approximations. 

We further explored the effects in our different model approximations using different levels of detailed network information. Specifically, our $\bar{R}$ model uses the observed numbers and lengths of node-independent shortest paths between each pair of nodes. In contrast, our $\tilde{R}$ models employ only the $p(k)$ degree distribution with a $p(\ell)$ path length distribution, and we considered differences using distributions of geodesic shortest paths versus node-independent shortest paths. 

Whereas our present approximations are more accurate at small temporal concurrency, productive future work might focus on the limiting behavior as $C\to 1$. Specifically, our approximation correctly captures $R=1$ at $C=1$, but the manner of approach as $C\to 1$ is noticeably incorrect compared to the simulated temporal network measurements, unless we artificially select an increased cohesion value as in Fig.~\ref{fig:result_ER_fit}. Given the relatively simple shape of the reachability versus concurrency curves, it is perhaps possible that a theory that is only correct in capturing the limiting $C\to 1$ behavior of reachability might be matched or otherwise combined with our present model to better approximate reachability over the whole $C\in [0,1]$ interval. Future studies might also explore the possible role of heterogeneity in actor-level concurrency across the network.

We believe the present study, focused on the role of \textit{temporal concurrency} and \textit{structural cohesion} in determining reachability, further emphasizes the need to better understand the interplay between the temporal and topological aspects in networks. 
With a more complete, integrated picture of this interplay, it may be possible in the future to identify different immunization strategies for outbreaks on empirical temporal networks in terms of their estimated structural and temporal properties. For example, such models could then be used to help predict possible benefits obtainable from targeting hub nodes in the underlying contact network versus individual-level or population-level interventions to decrease concurrency. 

\bigskip
\section*{Acknowledgements}
Research reported in this publication was supported by the Eunice Kennedy Shriver National Institute of Child Health \& Human Development of the National Institutes of Health under Award Number R01HD075712 and the James S. McDonnell Foundation 21st Century Science Initiative - Complex Systems Scholar Award grant \#220020315. The content is solely the responsibility of the authors and does not necessarily represent the official views of any sponsors.

\hfill

\appendix
\renewcommand\thefigure{\thesection.\arabic{figure}}    
\section{Construction of the exponential degree distribution networks}

\begin{figure}[!ht]
\centering
\includegraphics[width=0.95\linewidth]{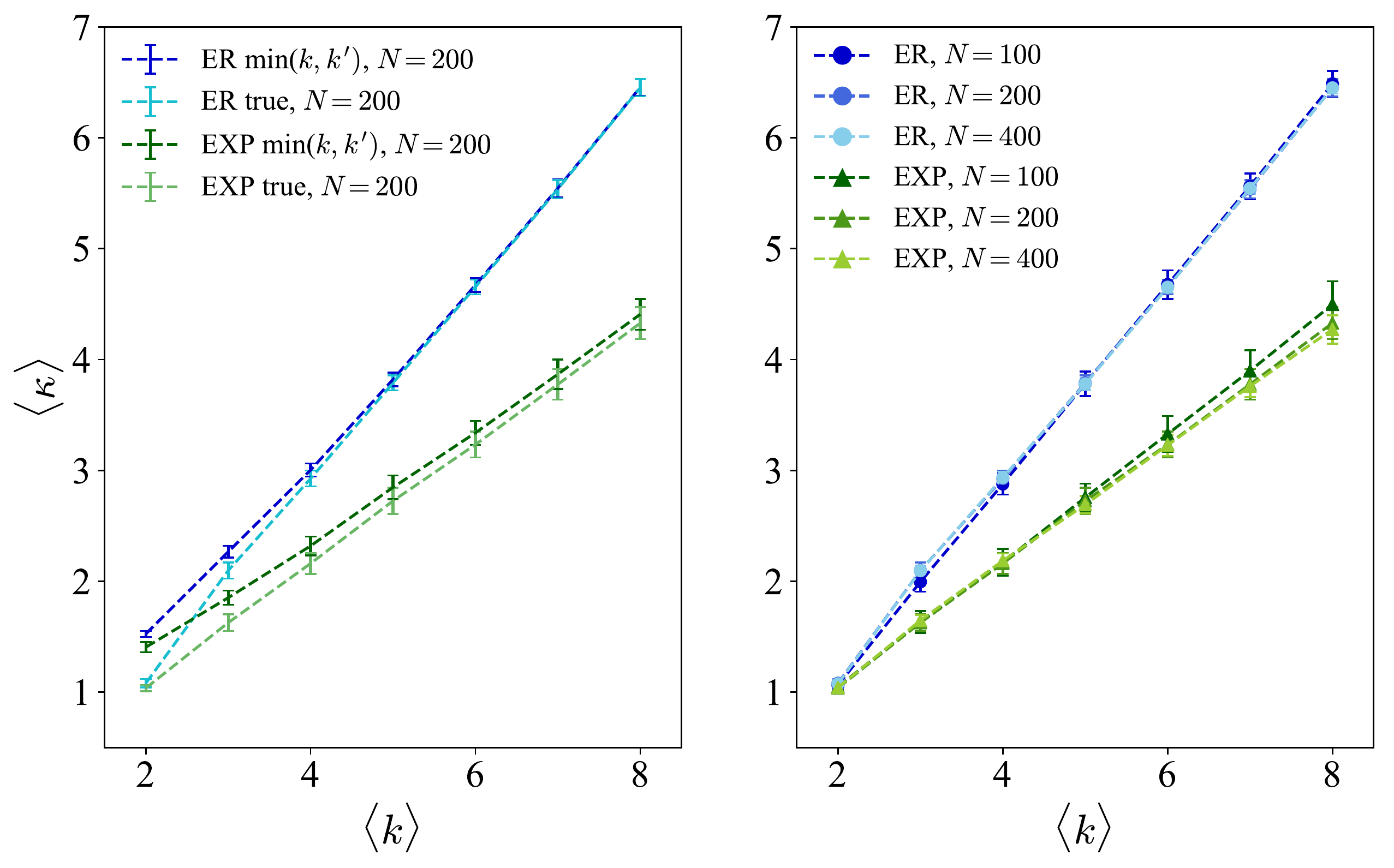}
\caption{Approximated and numerically estimated structural cohesion of ER and exponential degree distribution networks (``EXP") as a function of average degree $\langle k \rangle$. Solid lines show the approximated structural cohesion averaging $\kappa_{max}(i,j) = \mathrm{min}(k_i,k_j)$ over node pairs for ER (black) and EXP (grey) networks with $N=100$. Dashed lines are numerical estimations of the structural cohesion for ER (blue shades) and EXP (green shades) networks of different network sizes ($N=100$, $200$, $400$). Standard deviations are shown.}
\label{fig:S1_cohesion}
\end{figure}

To complement our tests on ER networks, we additionally consider networks with exponential degree distributions that have been rewired to match the structural cohesion of ER networks having the same mean degree. 
We construct a network with an exponential degree distribution using the {\tt configuration\_model} function in the NetworkX package in python, which follows steps described in~\cite{Newman2003}. A degree sequence $\{k_i\}$ for nodes $i=1,\cdots,N$ is generated by independent draws from the given  distribution $p(k)=\langle k \rangle^{-1} e^{{-k}/{\langle k \rangle}}$, where $\langle k\rangle$ is the desired mean degree. We used the largest connected component from the generated network. We removed self-loop and multi-edges and only accepted the resulting network if the mean degree was within $0.1$ of the desired $\langle k\rangle$. We note in particular that this procedure does not properly sample the space of simple configuration model graphs without self-loops and multi-edges~\cite{Dan2018Siam}. But for our present purposes of using these networks as random examples, we do not rely on obtaining a proper sampling of the space. We have not shown figures here exploring our approximations for these exponential degree distribution networks, since they are qualitatively similar to that discussed for ER networks in the main text, in particular having better accuracy at small $C$.

We note that the exponential degree distribution networks as generated to this point of the procedure have natural levels of structural cohesion that are different from ER networks with the same mean degree, as shown in Fig.~\ref{fig:S1_cohesion}. Because of the important role of structural cohesion in the present work, we seek to remove this difference between the exponential degree and ER networks. In Fig.~\ref{fig:S1_cohesion}(a), we see that the observed structural cohesion $\langle\kappa\rangle$ in these random graphs is very close to their upper bounds given by averaging over $\min(k_i,k_j)$, except at small mean degrees $\langle k\rangle$. In Fig.~\ref{fig:S1_cohesion}(b), we see that there is very little finite-size effect in the observed structural cohesion values on these graphs. (As an aside, we note that the empirical degree distributions in the largest connected component are generally slightly right-shifted from the imposed degree distribution before restriction to the largest connected component. This shift thereby increases the upper bound for structural cohesion obtained by averaging over $\min(k,k')$.)

To tune networks to a desired structural cohesion --- specifically, to make networks with ER and exponential degree distributions but with the same structural cohesion --- we rewire the links as follows (see, e.g., \cite{Maslov2002Specificity,Jo2014Generalized}). We randomly choose two links $(i,j)$ and $(iâ,jâ)$. If cutting these links does not break the network up into multiple components, we cut these links and then replace them with either $(i,iâ)$ and $(j,jâ)$ or $(i,jâ)$ and $(iâ,j)$. In so doing, we reject new candidate edges that generate multi-edges or self-loops and then select the pair of edges that make the new structural cohesion closest to the desired value. If neither rewiring option successfully moves the cohesion closer to the target value, the original cut edges are restored. By this method, the degree distribution remains constant while the degree-degree correlation and the structural cohesion change. We repeat this rewiring process until either the target value of structural cohesion is obtained (to within a tolerance here of $0.025$) or, if the target is not achieved within $E$ rewires then the process is restarted with a new random graph generated from the distribution.

\begin{figure}[!htp]
\centering
\includegraphics[width=0.8\linewidth]{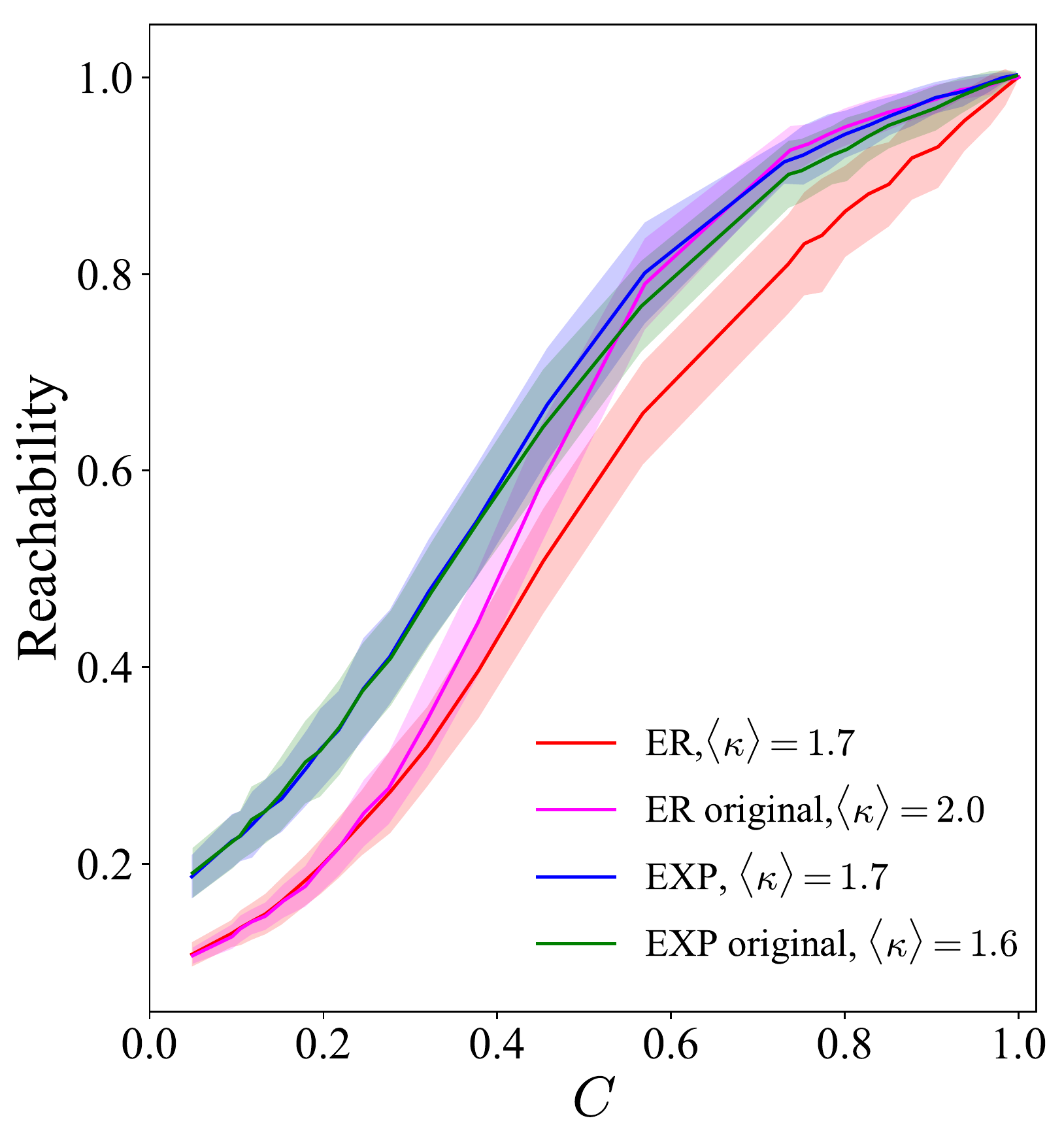}
\caption{Reachability $R$ versus temporal concurrency $C$ for random graphs with different degree distributions. The ``ER origin" graphs are Erd\H{o}s-R\'enyi graphs with mean degree $\langle k\rangle=3$, yielding an empirically expected cohesion $\langle\kappa\rangle=2$ (magenta line). The ``ER" graphs are with mean degree $\langle k\rangle=3$ rewired to obtain expected cohesion $\langle\kappa\rangle=1.7$ (red line). The ``EXP origin" graphs here are exponential degree distributions with the same $\langle k\rangle=3$, yielding an empirically expected cohesion $\langle \kappa \rangle = 1.6$ (green line). The ``EXP" denotes rewired graphs having exponential degree distribution, yielding expected cohesion $\langle\kappa\rangle=1.7$ as described in the Appendix. Filled area shows the standard deviation.}
\label{fig:erexp_control_cohesion}
\end{figure}

Figure~\ref{fig:erexp_control_cohesion} demonstrates the reachability of rewired ER and exponential degree distribution networks with the same average degree ($\langle k \rangle = 3$) and structural cohesion ($\langle \kappa \rangle = 1.7$). Even though the mean degree and structural cohesions of these random graphs are the same, the relationship between reachability and concurrency are noticeably different in the figure. This observation further motivates the development of our approximations in the main text in terms of degree distributions and path lengths.

\bibliographystyle{apsrev4-1}

\bibliography{main.bbl}
\end{document}